\documentclass[namedreferences]{solarphysics}

\usepackage[hyperref,optionalrh,showbiblabels]{spr-sola-addons} % For Solar Physics 
\usepackage{graphicx}        % For eps figures, newer & more powerfull
\usepackage{color}           % For color text: \color command
\usepackage{breakurl}        % For breaking URLs easily trough lines
\renewcommand{\vec}[1]{{\mathbfit #1}}

\newcommand{\aap}{    {\it Astron. Astrophys.}}
\newcommand{\aaps}{   {\it Astron. Astrophys. Suppl.}}

\newcommand{\aj}{     {\it Astron. J.}} 
\newcommand{\apj}{    {\it Astrophys. J.}}
\newcommand{\apjl}{   {\it Astrophys. J. Lett.}}

\newcommand{\mnras}{  {\it Mon. Not. Roy. Astron. Soc.}}

\newcommand{\solphys}{{\it Solar Phys.}}

\chardef\us=`\_

\begin{document}

\begin{article}
\begin{opening}

\title{Estimating solar flux density at low radio frequencies \\
using a sky brightness model}

\author[addressref=aff1,corref,email={div@ncra.tifr.res.in}]{\inits{D.}\fnm{Divya}~\lnm{Oberoi}}%\sep
\author[addressref=aff1,email={rohit@ncra.tifr.res.in}]{\inits{R.}\fnm{Rohit}~\lnm{Sharma}}%\sep
\author[addressref=aff2]{\inits{A.G}\fnm{Alan E.E.}~\lnm{Rogers}}%\sep
%\author{\inits{}\fnm{}~\lnm{}\orcid{}}
%\author{P.~\surname{Author-a}$^{1}$\sep
%        E.~\surname{Author-b}$^{1}$\sep
%        M.~\surname{Author-c}$^{2}$      
%       }

%   \institute{$^{1}$ First affiliation
%                     email: \url{e.mail-a} email: \url{e.mail-b}\\ 
%              $^{2}$ Second affiliation
%                     email: \url{e.mail-c} \\
%             }
\address[id=aff1]{National Centre for Radio Astrophysics, Tata Institute of Fundamental Research, Pune 411007, India}
\address[id=aff2]{MIT Haystack Observatory, Westford MA 01886, USA}

\runningauthor{Oberoi et al.}
\runningtitle{Estimating solar flux density at low radio frequencies using a sky brightness model}

\begin{abstract}
Sky models have been used in the past to calibrate individual low radio frequency telescopes.
Here we generalize this approach from a single antenna to a two element interferometer and formulate the problem in a manner to allow us to estimate the flux density of the Sun using the normalized cross-correlations (visibilities) measured on a low resolution interferometric baseline.
For wide field-of-view instruments, typically the case at low radio frequencies, this approach can provide robust absolute solar flux calibration for well characterized antennas and receiver systems.
It can provide a reliable and computationally lean method for extracting parameters of physical interest using a small fraction of the voluminous interferometric data, which can be prohibitingly compute intensive to calibrate and image using conventional approaches.
We demonstrate this technique by applying it to data from the Murchison Widefield Array and assess its reliability.  
\end{abstract}
\end{opening}
%-------------------------------------------------

\section{Introduction}
\label{Sec:intro}
Modern low radio frequency arrays use active elements to provide sky noise dominated signal over large bandwidths (e.g., LOFAR, LWA and MWA).
At these long wavelengths it is hard to build test setups to determine the absolute calibration for antennas or antenna arrays.
Models of the radio emission from the sky have, however, successfully been used to determine absolute calibration of active element arrays \citep{Rogers2004-cal-using-Gbg}.
Briefly, the idea is that the power output of an antenna, W, can be modeled as:
\begin{eqnarray}
\label{Eq:power-1}
%\nonumber
W(LST) = W_{i,i}(LST) = g_i g_i^*\ <V_i V_i^*>,
\end{eqnarray}
where LST is the local sidereal time; 
i is an index to label antennas;
$g_i$ is the instrumental gain of the $i^{th}$ antenna;
$^{*}$ represents complex conjugation;
$V_i$ is the voltage measured by the radio frequency probe 
and angular brackets denote averaging over time.
In temperature units, $ V_i V_i^*$ is itself given by:
\begin{equation}
\label{Eq:power-2}
V_{i,i} = V_i V_i^* =\frac{1}{2} A_{e} \Delta \nu \int_{\Omega} T_{Sky}(\vec{s})\ P_N(\vec{s})\ d\Omega + T_{Rec},
\end{equation}
where
$A_{e}$ is the effective collecting area of the antenna;
$\Delta \nu$ is the bandwidth over which the measurement is made; 
$\vec{s}$ is a direction vector in a coordinate system tied to the antenna (e.g., altitude-azimuth);
$T_{Sky}(\vec{s})$ is the sky brightness temperature distribution;
$P_N(\vec{s})$ is the normalized power pattern of the antenna;
$d\Omega$ represents the integration over the entire solid angle; and $T_{Rec}$ is the receiver noise temperature and also includes all other terrestrial contributions to the antenna noise.
The sky rotates above the antenna at the sidereal rate, hence the measured $W$ is a function of the local sidereal time.
The presence of strong large scale features in $T_{Sky}$ lead to a significant sidereal variation in $W$, even when averaged over the wide beams of the low radio frequency antennas.
If a model for $T_{sky}(\vec{s})$ is available at the frequency of observation and $P_N(\vec{s})$ is independently known, the integral in Eq. \ref{Eq:power-2} can be evaluated.
Assuming that either $g$ is stable over the period of observation, or that its variation can be independently calibrated, the only unknowns left in Eqs. \ref{Eq:power-1} and \ref{Eq:power-2} are $T_{Rec}$ and the instrumental gain $g$.
\citet{Rogers2004-cal-using-Gbg} successfully fitted the measurements with a model for the expected sidereal variation and determined both these free parameters.
This method requires observations spanning a large fraction of a sidereal day to be able to capture significant variation in $W$.
As $T_{sky}(\vec{s})$ does not include the Sun, which usually dominates the antenna temperature, $T_{Ant}$, at low radio frequencies, such observations tend to avoid the times when the Sun is above the horizon.

Our aim is to achieve absolute flux calibration for the Sun.
With this objective, we generalize the idea of using $T_{Sky}$ for calibrating an active antenna described above, from total power observations with a single element to a two element interferometer with well characterized active antenna elements and receiver systems.
Further, we pose the problem of calibration in a manner which allows us take advantage of the known antenna parameters to compute the solar flux using a sky model.
We demonstrate this technique on data from the Murchison Widefield Array (MWA) which operates in the 80--300 MHz band. 
A precursor to the Square Kilometre Array, the MWA is located in the radio quiet Western Australia.
Technical details about the MWA design are available in \citet{Lonsdale2009-MWA-design,Tingay2013-MWA-design}.
The MWA science case is summarized in \citet{Bowman2013-MWA-science} and includes solar, heliospheric and ionospheric studies among its key science focii.

With its densely sampled $u-v$ plane and the ability to provide spectroscopic imaging data at comparatively high time and spectral resolution, the MWA is very well suited for imaging the spectrally complex and dynamic solar emission \citep{Oberoi2011}.
Using the MWA imaging capabilities to capture the low level variations seen in the MWA solar data requires imaging at high time and frequency resolutions, 0.5 s and about hundred kHz, respectively.
In addition, the large number of the elements of the MWA (128), which give it good imaging capabilities, also lead to an intrinsically large data rate (about 1 TB/hour for the usual solar observing mode).
Hence, imaging large volumes of solar MWA data is challenging from perspectives ranging from data transport logistics, to computational and human resources needed.
Hence, a key motivation for this work was to develop a computationally inexpensive analysis technique capable of extracting physically interesting information from a small fraction of these data without requiring full interferometric imaging.
Such a technique also needs to be amenable to automation so that it can realistically be used to analyze large data-sets spanning thousands of hours.

The basis of this technique is formulated in Sec. \ref{Sec:formalism} and its implementation for the MWA data is described in Sec. \ref{Sec:implementation}.
The results and a study of the sources of random and systematic errors are presented in sections \ref{Sec:results} and \ref{Sec:errors}, respectively. 
A discussion is presented in Sec. \ref{Sec:discussion}, followed by the conclusions in Sec. \ref{Sec:conclusions}.

\section{Formalism}
\label{Sec:formalism}
The response of a baseline to the sky brightness distribution, $I(\vec{s})$, can be written as 
\begin{equation}
\label{Eq:baseline-response-1}
%%r = \frac{1}{2}\ A_{e}\ \Delta \nu\ \int_{\Omega} I(\vec{s})\ P_N(\vec{s})\ cos\frac{2 \pi \nu\ \vec{b} \cdot \vec{s}}{c}\ d\Omega,
V(\vec{b})  =  \frac{1}{2}\ A_{e}\ \Delta \nu \int_{\Omega} I(\vec{s})\ P_N(\vec{s})\ e^{\frac{-2 \pi i\ \nu \vec{b} \cdot \vec{s}}{c}}  d\Omega,
\end{equation}
where $V(\vec{b})$ is the measured cross correlation for the baseline $\vec{b}$;
$A_{e}$ is the effective collecting area of the antennas; 
$\Delta \nu$ is the bandwidth over which the measurement is made; 
$I(\vec{s})$ is the sky brightness distribution 
and 
$P_N(\vec{s})$ is the normalized antenna power pattern \citep{Synthesis-Imaging-1999}.
The antennas forming the baseline are assumed to be identical.
In terms of vector components this can be expressed as follows:
\begin{eqnarray}
\label{Eq:baseline-response}
\nonumber
V(u,v,w)  =  \frac{1}{2}\ A_{e}\ \Delta \nu \int \int I(l,m)\ P_N(l,m)\\
e^{-2 \pi i\ \{ul + vm + w(\sqrt{1 - l^2 - m^2}-1)\}}  \frac{dl\ dm }{\sqrt{1 - l^2 -m^2}},
\end{eqnarray}
where 
$u,v$ and $w$ are the components of the baseline vector $\vec{b}$, expressed in units of $\lambda$, in a right handed Cartesian coordinate system with $u$ pointing towards the local east, $v$ towards the local north and $w$ along the direction of the phase center,
and
$l,m$ and $n$ are the corresponding direction cosines with their origin at the phase center \citep{Synthesis-Imaging-1999}.
Compensation for the geometric delay between the signals arriving at the two ends of the baseline prior to their multiplication leads to the introduction of the minus one in the coefficient of $w$ in the exponential.
It is assumed that $\Delta \nu$ is narrow enough that variations of $P$ and $I$ with $\nu$ can be ignored.

We assume both the signal chains involved to also be identical.
In terms of the various sources of signal and noise contributing noise power to an interferometric measurement, the normalized cross-correlation coefficient, $r_N$, measured by a baseline can be written as
\begin{equation}
\label{Eq:r_N}
r_N = \frac{T_{\vec{b}}}{\overline{T_{Sky}} + T_{Rec} + T_{Pick-up}}.
\end{equation}
The numerator represents the signal power which is correlated between the two elements forming the baseline, $T_{\vec{b}}$, and the denominator is the sum of all the various contributions to the noise power of the individual elements.
$T_{Rec}$ represents the noise contribution of the signal chain, $T_{Pick-up}$ the noise power picked up from the ground and $\overline{T_{Sky}}$ the beam-averaged noise contribution from the sky visible to the antenna beam.
$\overline{T_{Sky}}$ is given by \citep{Synthesis-Imaging-1999}: 
\begin{equation}
\label{Eq:T-sky}
\overline{T_{Sky}} = \frac{1}{\Omega_P}\int_{\Omega} T_{Sky}(\vec{s})\ P_N(\vec{s})\ d\Omega.
\end{equation}
Here $\Omega_P$ is the solid angle of the normalized antenna beam, $P_N$.
The advantage of using a normalized quantity like $r_N$ is that, unlike Eq. \ref{Eq:power-1}, it is independent of instrumental gains, $g_i$s. 

For reasonably well characterized instruments, reliable estimates for $P(\vec{s})$, $T_{Rec}$ and $T_{Pick-up}$ are available from a mix of models and measurements.
Prior work by \citet{Rogers2004-cal-using-Gbg} has demonstrated that for antennas with wide fields of view, which allow for averaging over small angular scale variations, the 408 MHz all sky map by \citet{Haslam1982-408-map}, scaled using an appropriate spectral index, can be used as a suitable sky model.

As the sky model does not include the Sun, for solar observations $\overline{T_{Sky}}$ is modeled as the sum of contribution of the sky model as given in Eq. \ref{Eq:T-sky} and the beam-averaged contribution of the Sun, $\overline{T_{\odot, P}}$.
The angular size of radio Sun can be a large fraction of a degree. 
So for estimating solar flux, one needs baselines short enough that their angular resolution is much larger than the angular size of the Sun.
The best suited baselines are the ones which resolve out the bulk of the smooth large angular scale Galactic emission averages out over the wide field-of-view
%\footnote{A HPBW of $30^{\circ}$ is a good representative number for the MWA.} 
while retaining practically the entire solar emission.
In order to account appropriately for the sky model emission picked up by the baseline, $T_{\vec{b}\ Sky}$, the numerator of Eq. \ref{Eq:r_N} is modeled as
\begin{equation}
\label{Eq:T-baseline-total}
T_{\vec{b}} = \overline{T_{\odot, P}} + T_{\vec{b}\ Sky}.
\end{equation}
Given the geometry of the baseline, $T_{\vec{b}\ Sky}$ can be computed by incorporating the phase term reflecting the baseline response from Eq. \ref{Eq:baseline-response} in Eq. \ref{Eq:T-sky},
\begin{eqnarray}
\label{Eq:T-baseline-sky}
\nonumber
T_{\vec{b},Sky}  = | \frac{1}{\Omega_P} \int_{\Omega} T_{Sky}(\vec{s})\ P_N(\vec{s})\\
e^{-2 \pi i\ {(ul\ + vm\ + w(\sqrt{1-l^2-m^2}-1)}}\ d\Omega |.
\end{eqnarray}
Thus, for solar observations, Eq. \ref{Eq:r_N} can be written as
\begin{equation}
\label{Eq:r_N_Sun} 
r_{N, \odot} = \frac{\overline{T_{\odot, P}} + T_{\vec{b}\ Sky}}
{\overline{T_{Sky}} + \overline{T_{\odot, P}} + T_{Rec} + T_{Pick-up}}.
\end{equation}
The LHS of Eq. \ref{Eq:r_N_Sun} is the measured quantity and once $\overline{T_{Sky}}$ and $T_{\vec{b}\ Sky}$ are available from a model, the only remaining unknown on the RHS is $\overline{T_{\odot, P}}$.
Once $\overline{T_{\odot, P}}$ has been computed, the flux density of the Sun, $S_{\odot}$, is given by
\begin{equation}
\label{Eq:S_Sun} 
S_{\odot} = \frac{2\ k\ \overline{T_{\odot,P}}}{\lambda^2}\ \Omega_{P}.
\end{equation}
One can thus estimate $S_{\odot}$ using measurements from a single interferometric baseline from a wide field of view instrument.

Additionally, if the angular size of the Sun, $\Omega_{\odot}$, is independently known, the average brightness temperature of the Sun, $\overline{T_{\odot}}$, is given by
\begin{equation}
\label{Eq:T_Sun} 
\overline{T_{\odot}} = \overline{T_{\odot, P}}\ \frac{\Omega_{P}}{\Omega_{\odot}}.
\end{equation}

\section{Implementation}
\label{Sec:implementation}
To illustrate this approach we use data from the MWA taken on September 3, 2013 as a part of the solar observing proposal G0002 from 04:00:40 to 04:04:48.
%{\bf 
One GOES C1.3 class and 4 GOES B class flares were reported on this day. 
Six minor type III radio bursts and one minor type IV radio burst were also reported. 
Overall the level of activity reported on this day was classified as low by {\tt solarmonitor.org}.
%}

The MWA provides the flexibility to spread the observing bandwidth across the entire RF band in 24 pieces, each 1.28 MHz wide, providing a total observing bandwidth of 30.72 MHz.
These data were taken in the so called {\em picket-fence mode} where 12 groups of 2 contiguous coarse channels were distributed across the 80--300 MHz band in a roughly log-spaced manner.
Here we work with the 10 spectral bands at 100 MHz and above.
The time and frequency resolution of these data are $0.5\ s$ and $40\ kHz$, respectively.
Some of the spectral channels suffer from instrumental artifacts, and were not used in this study. 

For an interferometric baseline Eq. \ref{Eq:power-1} can be generalized to 
\begin{equation}
\label{Eq:power-baselines}
W_{i,j} = g_i g_j^* <V_i V_j^*>,
\end{equation}
where i and j are labels for the antennas comprising the baseline,
and the corresponding generalization for Eq. \ref{Eq:power-2} is given in Eqs. \ref{Eq:baseline-response-1} or \ref{Eq:baseline-response}. 
The LHS of Eq. \ref{Eq:r_N_Sun} is the measurable and is constructed as given below from the observed quantities:
\begin{equation}
\label{Eq:r_N_Sun-def} 
r_{N,\odot} = \frac{W_{i,j}}{\sqrt{W_{i,i} \times W_{j,j}}},
\end{equation}
It is evident from Eqs. \ref{Eq:power-baselines} and \ref{Eq:r_N_Sun-def} that $r_{N,\odot}$ is independent of the instrumental gain terms, making it suitable for the present application. 
In the following sub-sections we discuss the various terms in Eq. \ref{Eq:r_N_Sun} needed for estimating $\overline{T_{\odot, P}}$.

\subsection{$P_N(\vec{s})$ and $T_{Pick-up}$}
\begin{figure}
\centerline{\includegraphics[trim={7.3cm 0cm 0cm -0.5cm},clip,scale=.4]{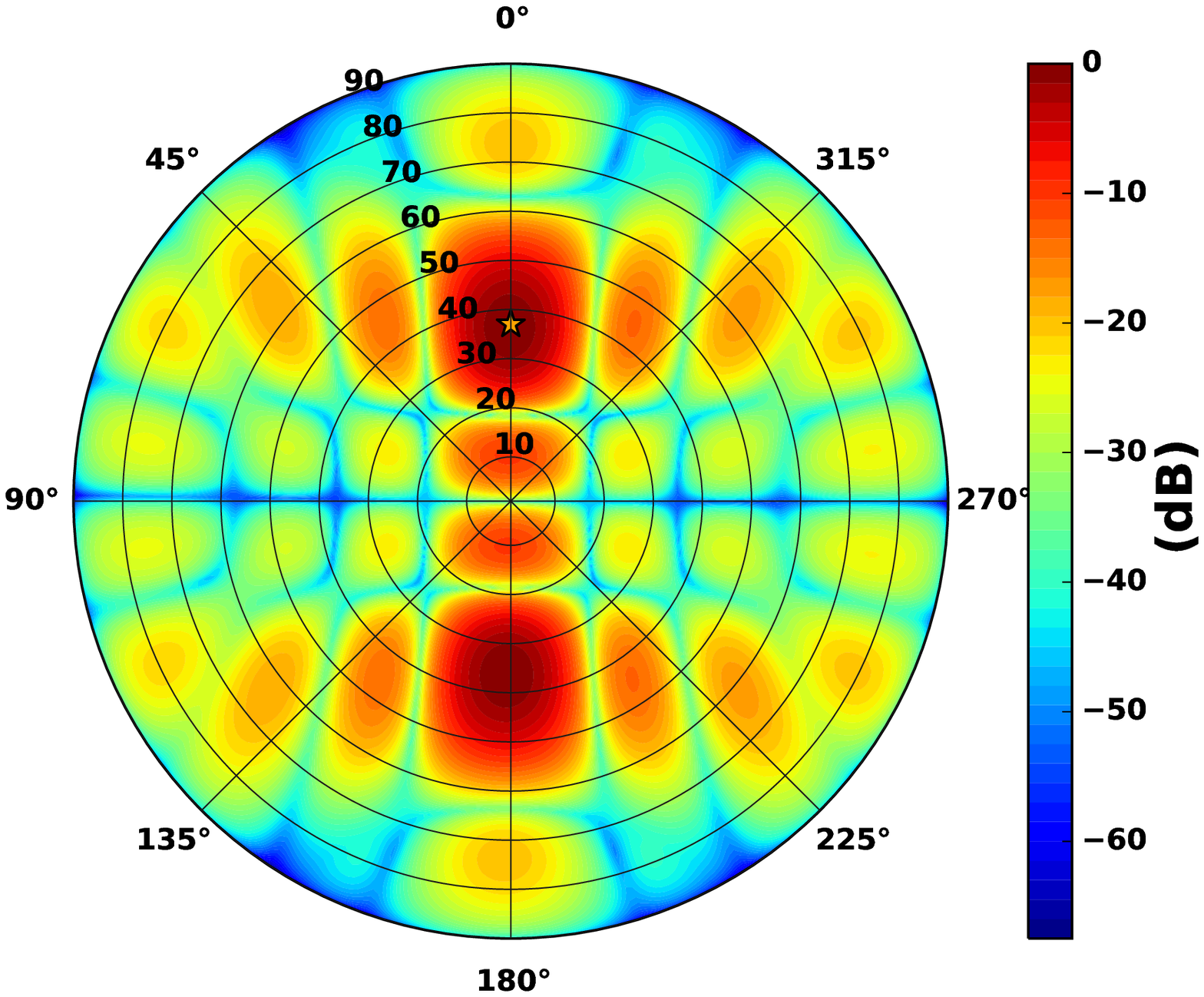}}
\centerline{\includegraphics[trim={7.5cm 0cm 0cm 1.0cm},clip,scale=.4]{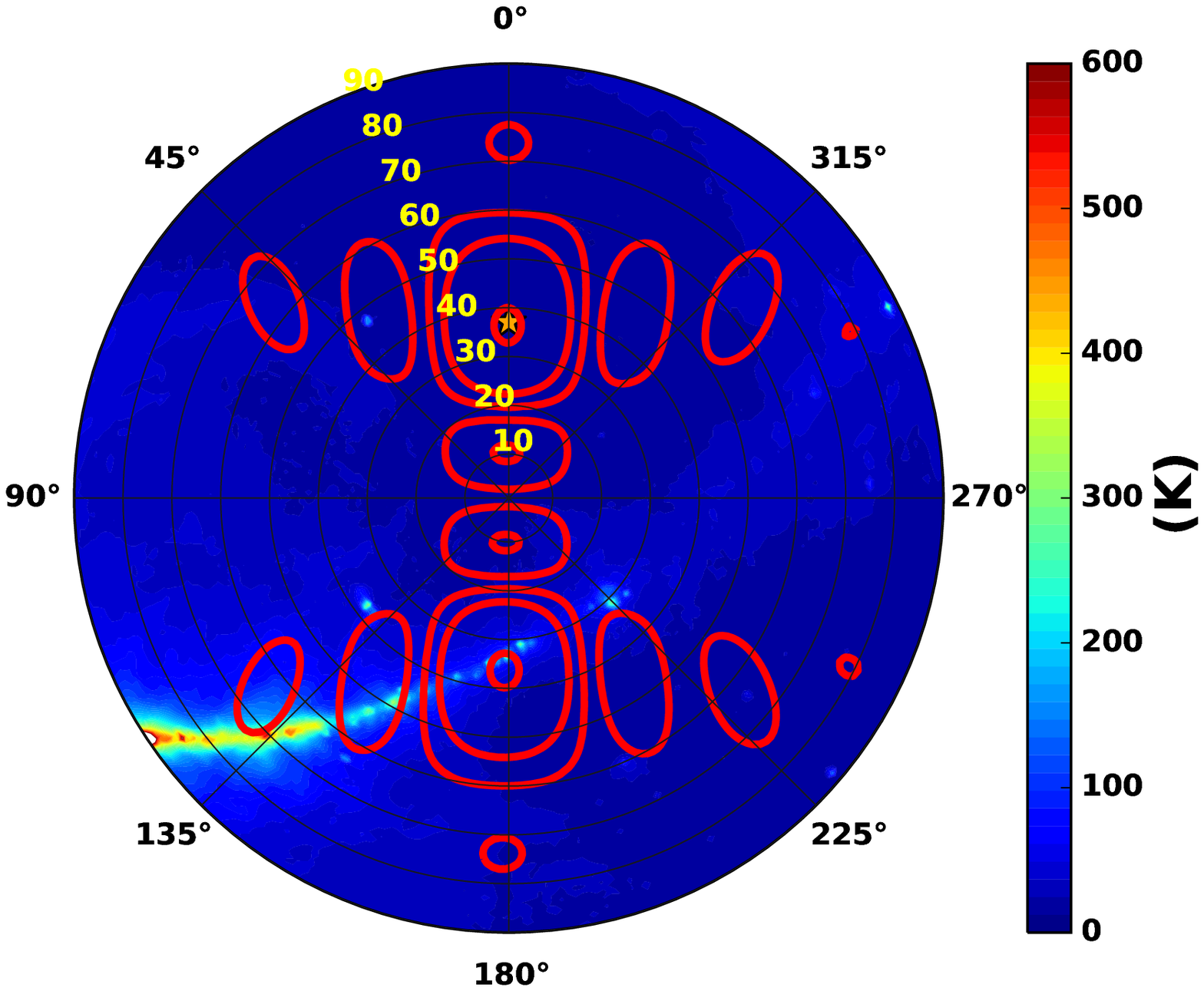}} 
\centerline{\includegraphics[trim={8.5cm 0cm 0cm 1.0cm},clip,scale=.4]{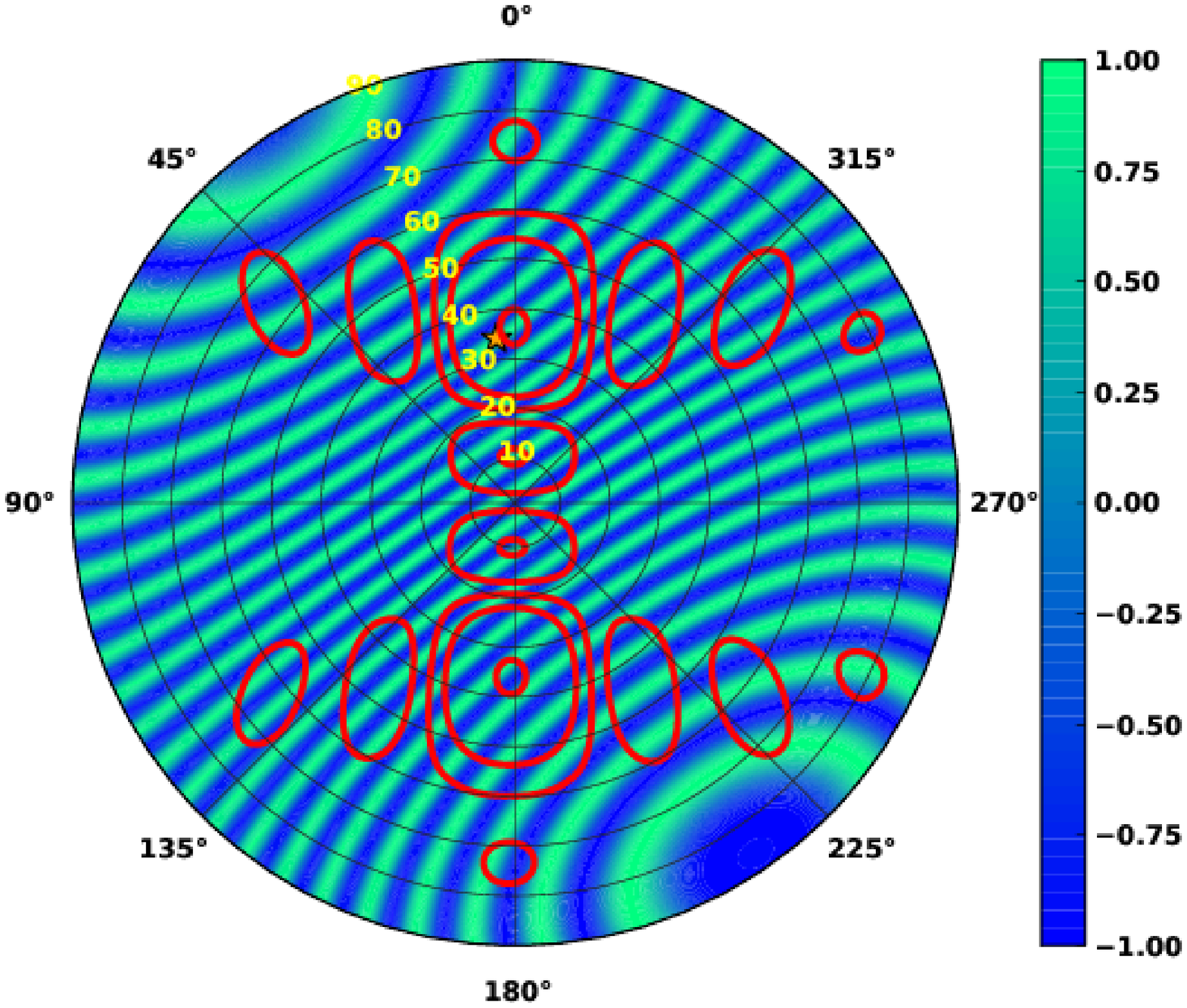}}
\caption{The top panel shows the model beam of a MWA tile in dB units at 238 MHz when pointed close to the direction of the Sun for the data presented here. 
The middle panel shows the all sky map at the same frequency, arrived at by scaling the \citet{Haslam1982-408-map} 408 MHz map with an $\alpha=-2.55$.
The bottom panel shows the cosine part of the phase term in the integrand in Eq. \ref{Eq:baseline-response} for baseline Tile011-Tile022 (u=7.86, v=6.19, w=4.56) at 238 MHz.
All the panels show the sky above the horizon in the altitude-azimuth coordinate system. 
Azimuth of 0$^{\circ}$ points towards north, the local zenith is at the center and the radial axis shows zenith angle in degrees. 
The location of the Sun for the mid point of the observing period is marked on all the panels.
The contours show the MWA beam from the top panel at values of 90\%, 9\% and 0.9\%. 
}
\label{Fig:1}
\end{figure}

Detailed electromagnetic simulations of the MWA antenna elements, referred to as tiles, including the effects of mutual coupling and finite ground screen, have been done to compute reliable models for $P_N(\vec{s})$ in the 100--300 MHz band. 
These simulations compute the {\em embedded patterns} using a numerical electromagnetic code (FEKO).
The MWA tiles comprise 16 dual-polarization active elements arranged in a $4\times4$ grid placed on a $5m \times 5m$ ground screen.
For efficiency of computing, we assume that a tile has only 4 distinct embedded patterns from which all 16 can be obtained by rotation and mirror reflection.
The full geometry is placed on welded wire ground screen over a dielectric earth. 
These simulations also allow us to compute the ground loss.
%A full $4\pi$ pattern is computed so that the ground loss can be estimated from the integrated power fraction in the lower hemisphere.
The beam pattern for a given direction is the vector sum of the embedded patterns for each of the 16 elements using the appropriate geometric phase delays.
The embedded patterns change slowly with frequency and we compute and store the real and imaginary parts for each polarization every 10 MHz at a $1^{\circ}\times 1^{\circ}$ azimuth-elevation grid for all the 4 embedded patterns.
Using these embedded pattern files, we can interpolate to compute the beam patterns for any given frequency and direction.
We also determine the ground loss as a function of frequency in terms of noise power which will be added to the receiver and sky noise.
This contribution to noise power, referred to as $T_{Pick-up}$, varies between 10--20 K.
An example MWA beam pattern at 238 MHz is shown in the top panel of Fig. \ref{Fig:1} for an azimuth and zenith angle of 0.0$^{\circ}$ and 36.4$^{\circ}$, respectively. 

\subsection{$T_{Rec}$}
The $T_{Rec}$ for the MWA has been modeled based on the radio frequency design and signal chain, and successfully verified against field measurements.
Though all MWA tiles are identical in design, they lie at differing distances from the receiver units where the data is digitized.
A few different flavors of cables are used to connect them to the receivers. 
The value of $T_{Rec}$ for a tile depends on the length and characteristics of this cable.
Here we have chosen to work only with the tiles using the shortest cable runs (90 m), which also give the best $T_{Rec}$ performance.
For these tiles the $T_{Rec}$ is close to 35 K at 100 MHz, drops gradually to about 20 K at 180 MHz and then increases smoothly to about 30 K at 300 MHz.

\subsection{Choice of sky model and spectral index}
The \citet{Haslam1982-408-map} all-sky map at 408 MHz, with an angular resolution of $0.^{\circ}85$, zero level offset estimated to be better than 2 K and random temperature errors on final maps $<$ 0.5 K \citep{Haslam1981}, is the best suited map for our application.
Its reliability has been independently established in prior work \citep{Rogers2004-cal-using-Gbg} and it is routinely used as the sky model at low radio frequencies.
A spectral index is used to translate the map to the frequency of interest.
The observed radio emission comes from both Galactic and extra-galactic sources and it is commonly assumed that the emission spectrum, averaged over sufficiently large patches in the sky, can be described simply by a spectral index, $\alpha$, typically defined in temperature units as $T \propto \nu^{\alpha}$.
The $\alpha$ can vary from one part of the sky to another so, in principle, one needs an all-sky $\alpha$ map to account for its variation across the sky. 
In practice, when averaged over the large angular scales corresponding to fields-of-view of low frequency elements (order $10^3\ deg^2$), $\alpha$ converges to a rather stable value.
There have been a few independent estimates of the spectral index of the Galactic background radiation and its variation as a function of direction \citep{Lawson1987-GB-spectral-index,Rogers2008-Tsky-spectral-index, Guzman2011-Tsky-spectral-index}.
The most recent of these studies computed a spectral index between 45 MHz and 408 MHz. 
It concluded that over most of the sky the spectral index is between 2.5 and 2.6, which is reduced by thermal absorption in much of the $|b| < 10^{\circ}$ region to values between 2.1 and 2.5.
This study also provided a spectral index map.
Here we work with only one pointing direction which is chosen to avoid the Galactic plane and use $\alpha=-2.55$, which is appropriate for this direction.
The middle panel of Fig. \ref{Fig:1} shows the model sky at 238 MHz derived from the \citet{Haslam1982-408-map} 408 MHz map.

\subsection{Computing $\overline{T_{Sky}}$ and $T_{\vec{b},\ Sky}$}
Once $T_{Sky}(\vec{s})$ and $P_N(\vec{s})$ are known, $\overline{T_{Sky}}$ can be computed using Eq. \ref{Eq:T-sky}.
Computing $T_{\vec{b},\ Sky}$ requires choosing a baseline.
The integrand in Eq. \ref{Eq:baseline-response} includes a phase term which is responsible for a given baseline averaging out the spatially smooth part of the emission in the beam.
For this application, the ideal baselines are the ones which are short enough for a source approaching $1^{\circ}$ to appear like an unresolved point source (Sec. \ref{subsec:MWA-pointing}), while the contribution of the smoothly varying Galactic emission drops dramatically as it gets averaged over multiple fringes of the phase term in Eq. \ref{Eq:baseline-response}. 
The heavily centrally condensed MWA array configuration provides many suitable baselines.
The cosine part of the phase term mentioned above is shown in the bottom panel of Fig. \ref{Fig:1} for an example baseline. 
Table \ref{Tab:1} lists the $T_{\vec{b},\ Sky}$ for all the different frequencies considered here for this baseline.

\subsection{Choice of angular size of Sun}
\label{subsec:size-of-sun}
While $S_{\odot}$ can be computed unambiguously in this formalism, computing $\overline{T_{\odot}}$ requires an additional piece of information, $\Omega_{\odot}$ (Eq. \ref{Eq:T_Sun}).
The MWA data can provide the images from which $\Omega_{\odot}$ can, in principle, be measured. 
Deceptively, however, this involves some complications.
Given the imaging dynamic range and the resolution of the MWA, the Sun usually appears as an asymmetric source with a somewhat complicated morphology.
This rules out the approach of fitting elliptical Gaussians to estimate the radio size of the Sun used in some of the earlier work \citep[e.g.][]{McLean-Sheridan-1985}.
Associating an angular size with such a source requires one to define a threshold and integrate the region enclosed within this contour to give the angular size of the Sun. 
The solar emission at these frequencies comes from the corona and this emission does not have a sharp boundary. 
The choice of the threshold is, hence, bound to be somewhat subjective.
Also, as mentioned in Sec. \ref{Sec:intro}, a key objective of this work is to develop a technique which is numerically much less intensive than interferometric imaging, so we cannot expect these solar radio images to be available.

In absence of more detailed information, our best recourse is to assume the Sun to be effectively a circular disc with a frequency dependent size given by the following empirical relationship (Stephen White; private communication):
\begin{equation}
\theta_{\odot} = 32.0 + 2.22 \times \nu_{GHz} ^{-0.60},
\label{Eq:Sun_size}
\end{equation}
where $\theta_{\odot}$ is the expected effective solar diameter in arcmin and $\nu_{GHz}$, the observing frequency in GHz.
%{\bf 
The solar radio images are well known to have non-circular appearance and their equatorial and polar diameters can be different by as much as 30\% at metre wavelengths \citep[e.g.][]{Lantos1992,Mercier2012}.
$\theta_{\odot}$ represents an {\em effective diameter} yielding the same surface area as the true solar brightness distribution.
%}
Values of $\theta_{\odot}$ are tabulated in Table \ref{Tab:2}.
While the solid angle subtended by the Sun is expected to vary with the presence of coronal features like streamers and coronal holes, and the phase of the solar cycle, this expected variation is only a fraction of its mean angular size.
Additionally, we note that the active emissions are usually expected to come from compact sources, so in presence of solar activity, this leads to a large underestimate of the true $T_{\odot}$ for active regions.
In spite of these limitations this approach provides very useful estimates of $\overline{T_{\odot}}$, especially for the quiet Sun.

\subsection{Choice of MWA pointing direction}
\label{subsec:MWA-pointing}
\begin{figure}
\centerline{\includegraphics[scale=.42,trim={0cm 0cm 0cm 1cm},clip]{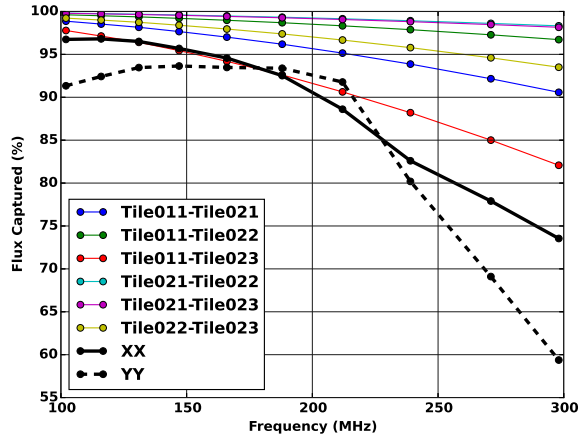}}
\caption{The figure shows the effects which need to be corrected for to obtain the true value of $\overline{S_{\odot}}$.
The black curves show the drop in  $P_{N}$, due to the Sun not being at the center of the MWA tile beams, as a function of frequency for X and Y polarizations.
The fraction of solar flux recovered by a few of the baselines used here are also shown as a function of frequency.
}
\label{Fig:2}
\end{figure}

The MWA tiles are pointed towards the chosen direction in the sky by introducing appropriate delays between the signals from the dipoles comprising a tile. 
These delays are implemented by switching in one or more of five independent delay lines, which provide delays in steps of two, for each of the dipoles \citep{Tingay2013-MWA-design}. 
A consequence of this discreteness in the delay settings is that all the different signals can be delayed by exactly the required amounts only for certain specific directions.
These directions are referred to as {\em sweet spots} and the MWA beams are expected to be closest to the modeled values towards these directions.
For this reason, for solar observations we point to the sweet spot nearest to the Sun, rather than the Sun itself.
For the data presented here, the nearest sweet spot was located at a distance of 4.18$^{\circ}$ from the Sun and implies that $P_{N}$ is not unity towards the direction of the Sun. 
Figure \ref{Fig:2} plots the value of $P_{N}$ towards the direction of the Sun as a function of frequency for both the polarizations.
The pointing center was also used as the phase center for computing the cross-correlations. 
We account carefully for the loss of flux on the short baselines due to this, using our chosen model for solar radio emission (Sec. \ref{subsec:size-of-sun}). 
The amplitude of the integral over the phase term in  Eq. \ref{Eq:baseline-response} over a circular disc of size given by Eq. \ref{Eq:Sun_size} located with an appropriate offset with respect to the phase center measures the solar flux picked up by a given baseline.
This quantity is also shown in Fig. \ref{Fig:2} as a fraction of the flux recovered by some example baselines as a function of frequency.
Both of these effects are corrected for in all subsequent analysis. 

\section{Results}
\label{Sec:results}
\begin{figure}
\centerline{\includegraphics[scale=.5,trim={0.2cm 0cm 0cm 1cm},clip]{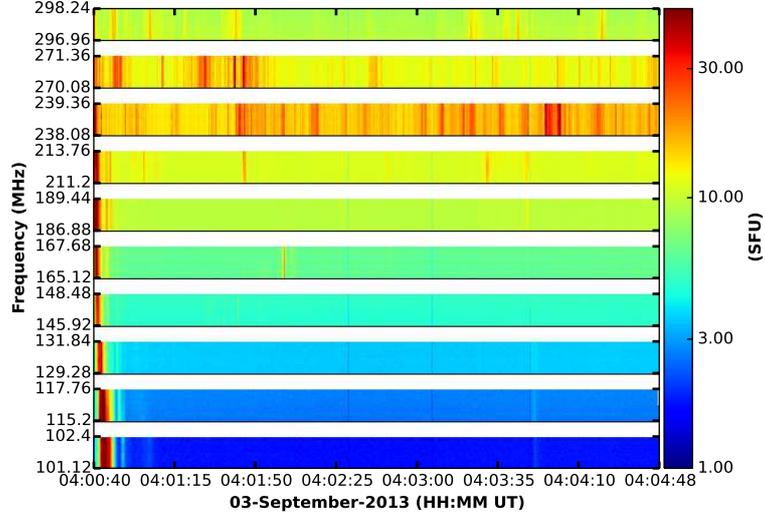}}
\caption{The computed values of $\overline{S_{\odot}}$ in units of SFU (1 SFU = 10$^4$ Jy) as a function of time for each of the 10 spectral bands.
A type-III like drifting feature is seen at the start of the data.
The start and stop frequencies of each of the bands are mentioned.
The individual bands usually span close to 2.5 MHz, with a few exceptions where the data was discarded due to instrument related issues or persistent satellite based radio frequency interference.
}
\label{Fig:3} % Fig 3
\end{figure}

\begin{figure}
\centerline{\includegraphics[scale=.4,trim={0.8cm 0cm 0cm 1cm},clip]{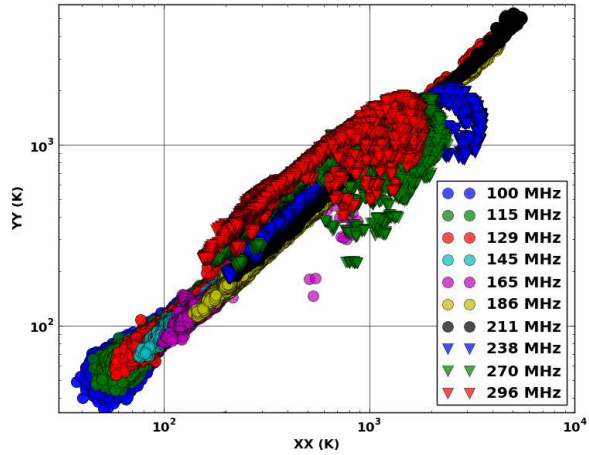}}
\caption{The X and Y axes show the values of $\overline{T_{\odot,P}}$ computed for the X and the Y polarizations, respectively, for one of the baselines.
A log scale has been used to emphasize the large range of $\overline{T_{\odot,P}}$ observed.
$\overline{T_{\odot,P}}$ corresponding to the quiescent emission spans a range of about an order of magnitude from about 50--500 K (Table \ref{Tab:1}), and the active emissions extend the range by another order of magnitude.
The measurements between the two polarizations are very consistent.
}
\label{Fig:pol_comparison} % Fig 4
\end{figure}

To provide an estimate of the magnitudes of the different terms in Eq. \ref{Eq:r_N_Sun}, Table \ref{Tab:1} lists representative values of these terms for all the ten observing bands spanning 100--300 MHz for the XX polarization for the baseline Tile011-Tile022.
Figure \ref{Fig:3} shows the dynamic spectra for $\overline{S_{\odot}}$ for the bands listed in Table \ref{Tab:1}.
As the bulk of the emission at these radio frequencies is thermal emission from the million K coronal plasma, the broadband featureless emission is not expected to have significant linear polarization.
The consistency between $\overline{T_{\odot,P}}$ computed for the XX and the YY polarizations, for all the ten spectral bands, is demonstrated in Fig \ref{Fig:pol_comparison}.

\begin{figure}
\centerline{\includegraphics[scale=.42,trim={0.0cm 0cm 0cm 1cm},clip]{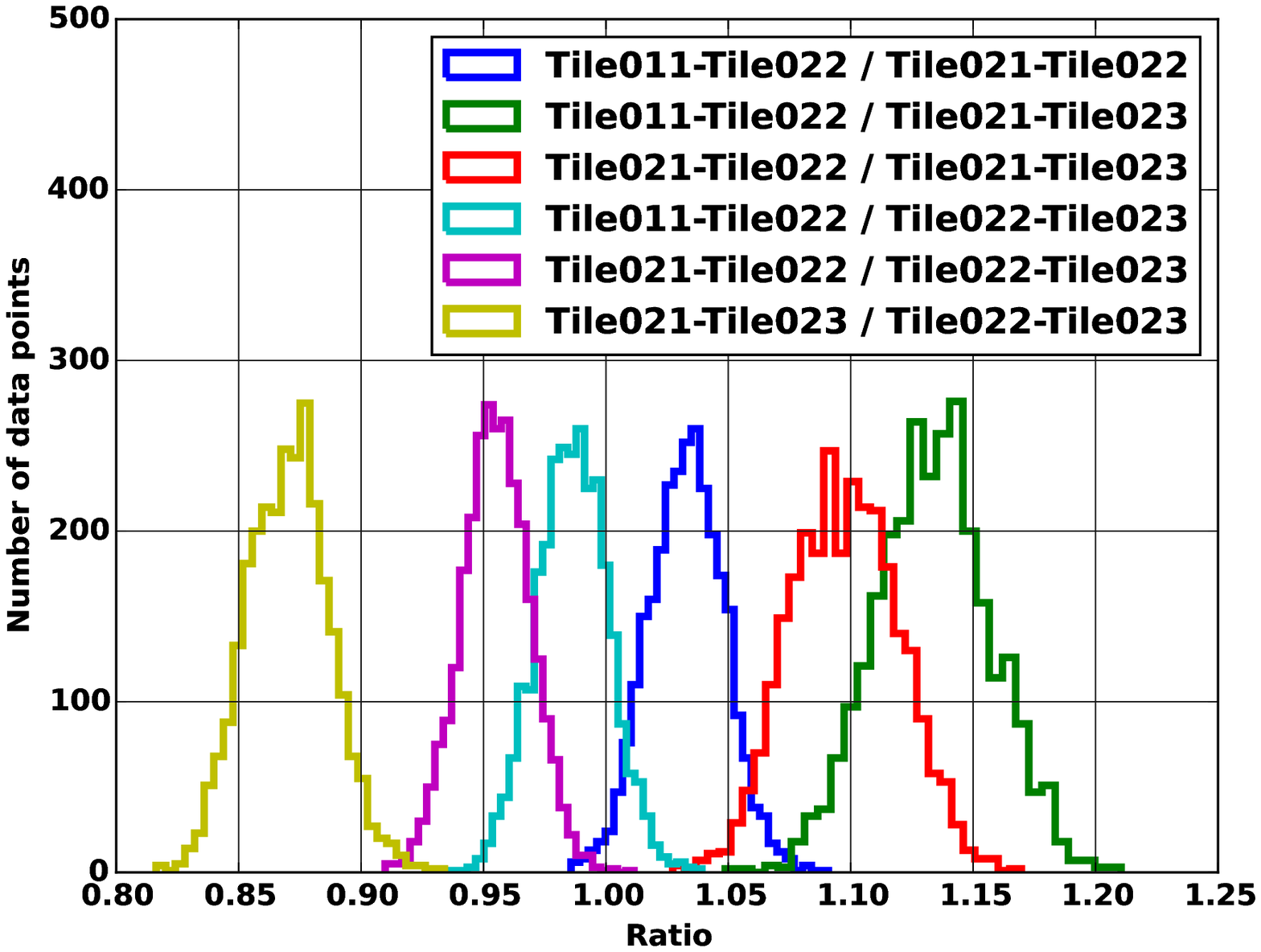}}
\caption{Histograms of the ratio of $\overline{T_{\odot,P}}$ measured using visibilities of the same polarization corresponding to the same spectral and time slice but different baselines, for the 167 MHz band.
}
\label{Fig:baseline_comparison1} % Fig 5
\end{figure}

\begin{figure}
\centerline{\includegraphics[scale=.43,trim={0.8cm 0cm 0cm 1cm},clip]{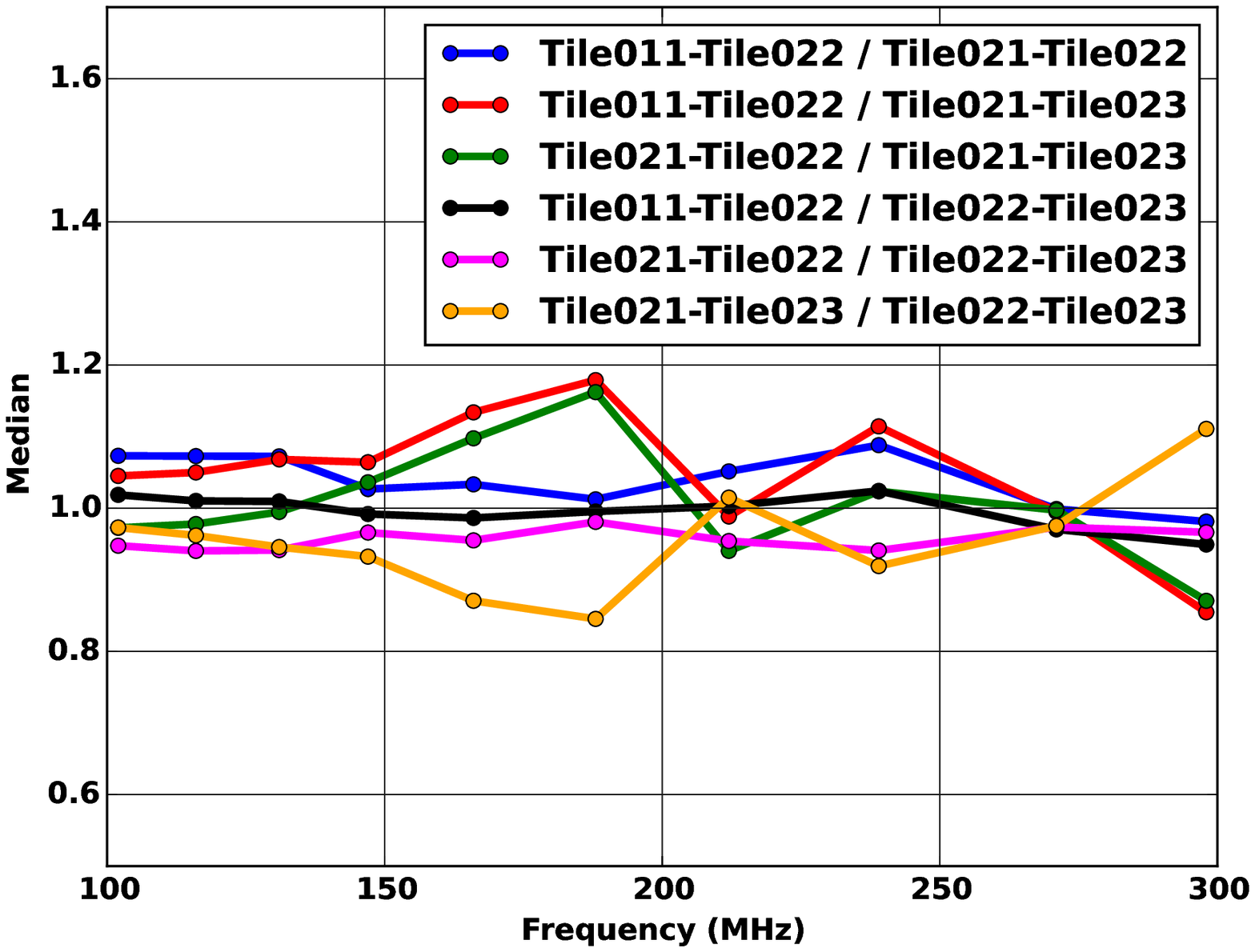}}
\centerline{\includegraphics[scale=.43,trim={0.8cm 0cm 0cm 1cm},clip]{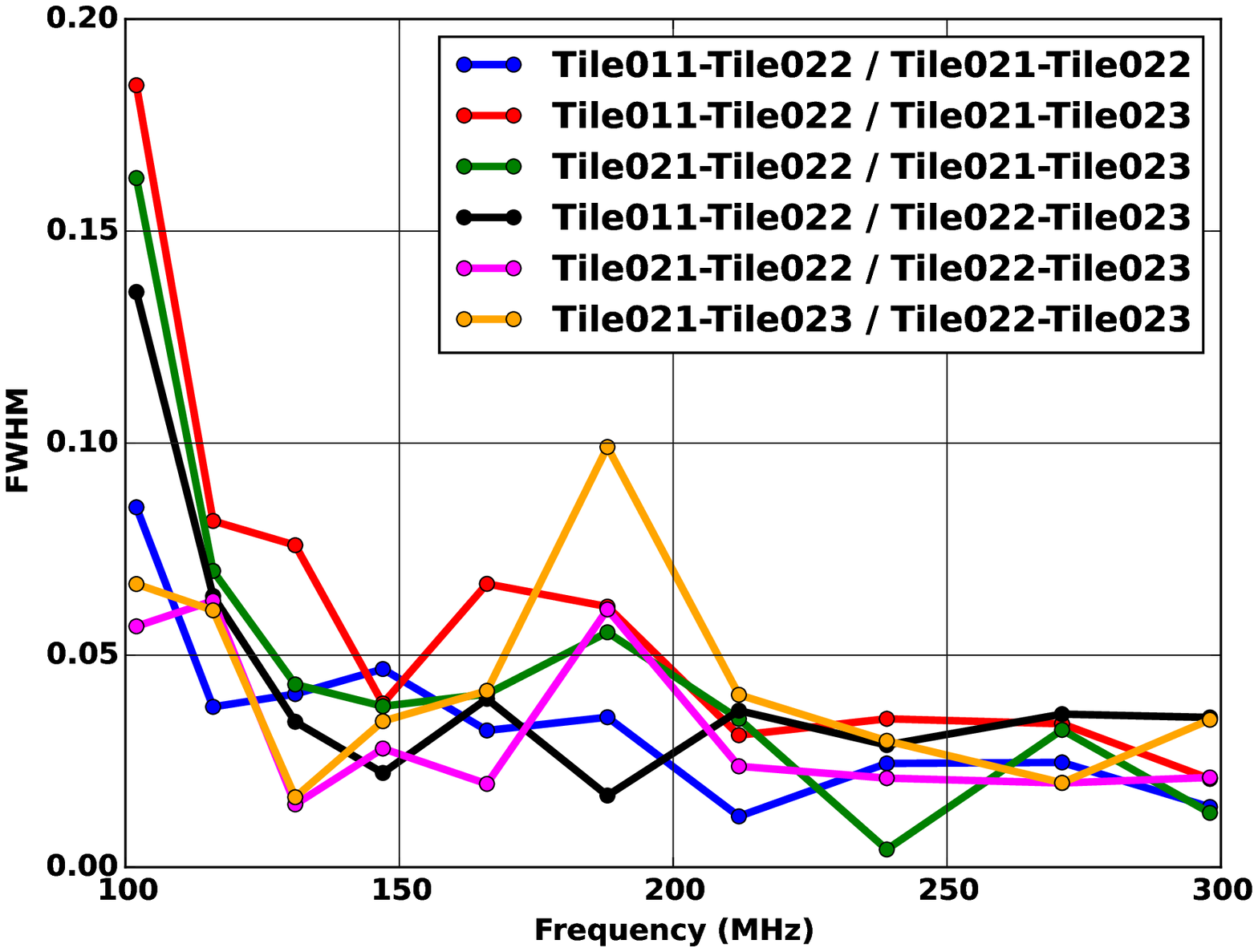}}
\caption{
The top panel shows the medians of the kind of histograms shown in Fig. \ref{Fig:baseline_comparison1} for all six baselines formed  between tiles Tile011, Tile021, Tile022 and Tile023 as a function of frequency. 
The bottom panel shows the FWHM of the same histograms.
}
\label{Fig:baseline_comparison2} % Fig 6
\end{figure}

The availability of many MWA baselines of suitably short lengths provides a convenient way to check for consistency between estimates of $\overline{T_{\odot,P}}$ from different baselines.
Here, we consider all six baselines formed between the following four tiles -- Tile011, Tile021, Tile022 and Tile023. 
Table \ref{Tab:2} shows the mean of the various parameters of interest over these six baselines, and RMS of these values computed over these baselines.
Figure \ref{Fig:baseline_comparison1} shows a comparison of the $\overline{T_{\odot,P}}$ computed using the data for the same polarization (XX) for these baselines.
The variations in the median values of these histograms of ratios of $\overline{T_{\odot,P}}$ measured on the same baselines are shown as a function of frequency in Fig \ref{Fig:baseline_comparison2}, along with the FWHM of the these histograms.

\section{Uncertainty estimates}
\label{Sec:errors}
\begin{figure}
\centerline{\includegraphics[scale=.51,trim={0cm 0cm 0cm 0cm},clip]{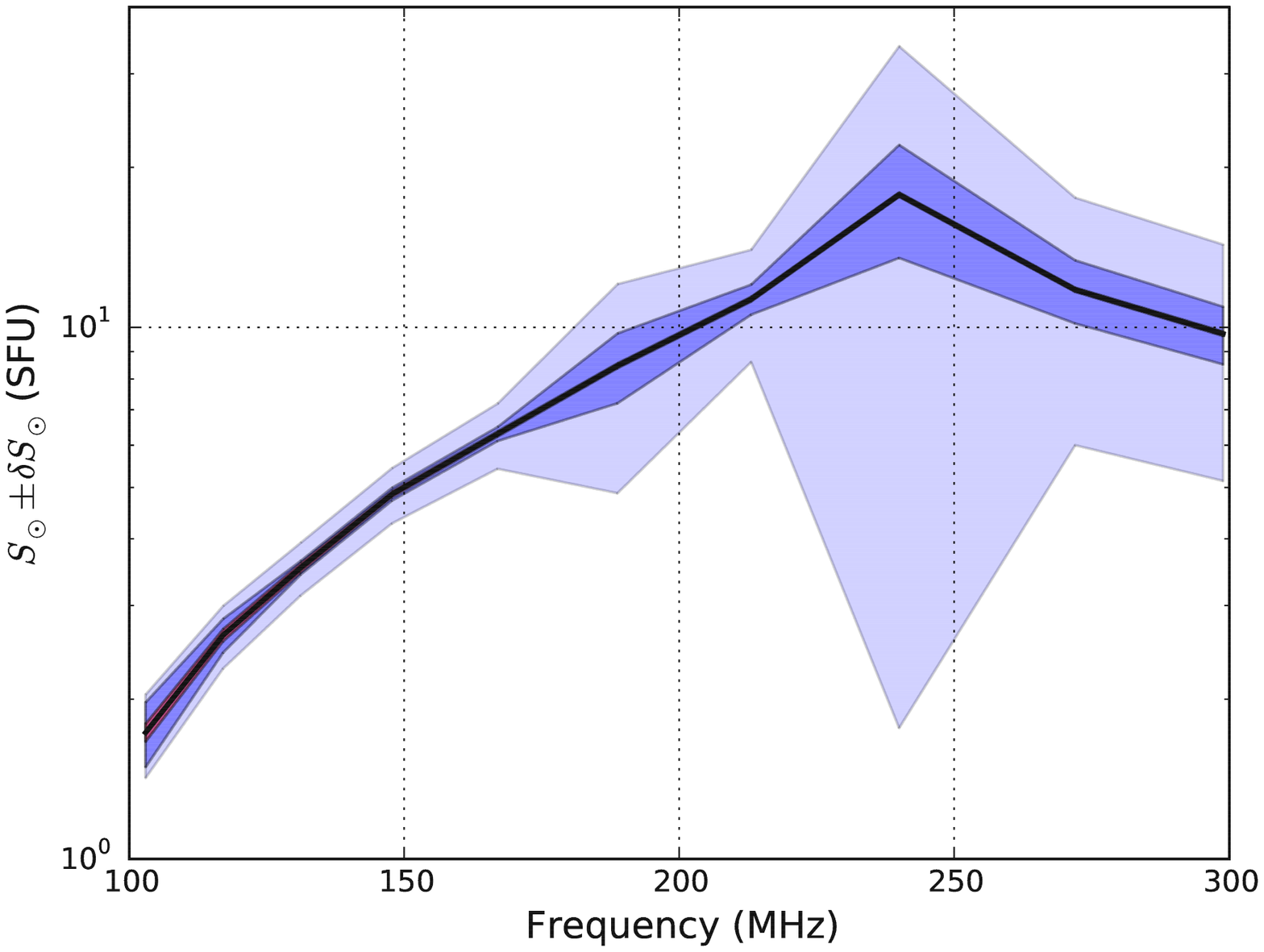}}
\caption{The measured values of mean $S_{\odot}$ from Table \ref{Tab:2} are shown by the central black line.
The red shaded region around it, only barely visible at the lowest frequencies, shows the 1$\sigma$ uncertainty due to theoretical thermal noise, $\delta S_{\odot,Th}$, which is $<$1\% of $S_{\odot}$ at higher frequencies.
The darker blue shaded region shows the 1$\sigma$ uncertainty due to the observed RMS, $\delta S_{\odot,Obs}$ from Table \ref{Tab:1}.
The light blue region shows the 1 $\sigma$ uncertainty in absolute value of $S_{\odot}$ after taking into account the known sources of systematic errors, $\delta S_{\odot,Abs}$.
}
\label{Fig:dS_vs_freq}
\end{figure}

The key quantity of physical interest is $S_{\odot}$. 
The intrinsic measurement uncertainty in $S_{\odot}$ due to thermal noise is given by
\begin{equation}
\delta S_{\odot,Th} = \frac{2\ k}{A_{eff}} \frac{T_{Sys}}{\sqrt{\Delta \nu\ \Delta t}},
\label{Eq:Thermal_uncertainty}
\end{equation}
where $T_{Sys}$, the system temperature, is the sum of all the terms in the denominator of Eq. \ref{Eq:r_N_Sun}, $A_{eff}$ is the effective collecting area of an MWA tile in m$^2$ (given by $\lambda^2/\Omega_P$), and $\Delta \nu$ and $\Delta t$ the bandwidth and the durations of individual measurements, respectively.
For the data presented here, $\delta S_{\odot,Th}$ lies in the range 0.02--0.06 SFU (Table \ref{Tab:1}). 
The uncertainty in $S_{\odot}$ due to thermal noise is at most a few \% and usually $<$1\%.
Figure \ref{Fig:dS_vs_freq} shows $S_{\odot}$, $\delta S_{\odot,Th}$ and $\delta S_{\odot,Obs}$, the observed RMS on $S_{\odot}$, as a function of $\nu$.
$\delta S_{\odot,Obs}$ exceeds $\delta S_{\odot,Th}$ by factors ranging from a few to almost two orders of magnitude, even during a relatively quiet times.
This establishes that $\delta S_{\odot,Obs}$ is dominated by intrinsic changes in $S_{\odot}$ and demonstrates the sensitivity of these observations to low level changes in $S_{\odot}$.

In addition to the random errors discussed above, estimates of $S_{\odot}$ will also suffer from systematic errors.
In fact, the uncertainty in the estimate of $S_{\odot}$, $\delta S_{\odot}$, is expected to be dominated by these systematic errors.
Following the usual principles of propagation of error and assuming the different sources of errors to be independent and uncorrelated, we estimate $\delta S_{\odot}$ considering the various known sources of error.
Rearranging Eq. \ref{Eq:r_N_Sun}, the primary measurable, $\overline{T_{\odot,P}}$, can be expressed as:
\begin{equation}
\label{Eq:T_Sun_P}
\overline{T_{\odot,P}} = \frac{r_{N, \odot}(\overline{T_{Sky}} + T_{Rec} + T_{Pick-up}) - T_{\vec{b}, Sky}}{1 - r_{N, \odot}},
\end{equation}
and $\delta\overline{T_{\odot,P,Abs}}$ the absolute error in, $\overline{T_{\odot,P}}$, is given by:
\begin{eqnarray}
\label{Eq:delta_T_Sun_P}
\nonumber
\delta\overline{T_{\odot,P,Abs}}^2 = \\
\nonumber
\delta r_{N,\odot}^2 \left(\frac{\overline{T_{Sky}}+T_{Rec}+T_{Pick-up}-T_{\vec{b},Sky}} {(1-r_{N,\odot})^2} \right)^2\\
\nonumber
+ \left( \delta \overline{T_{Sky}}^2 + \delta T_{Rec}^2 + \delta T_{Pick-up}^2 \right) \left( \frac{r_{N,\odot}}{1-r_{N,\odot}} \right)^2\\
+ \delta T_{\vec{b}, Sky}^2 \left( \frac{1}{1-r_{N,\odot}} \right)^2,
\label{Eq:Systematic_uncertainty}
\end{eqnarray}
where the pre-fix $\delta$ indicates the error in that quantity.
$\delta S_{\odot,Abs}$ can be computed from $\delta \overline{T_{\odot,P,Abs}}$ using an equation similar to Eq. \ref{Eq:S_Sun}, though it is more convenient to discuss the different error contributions in temperature units.

For the MWA tiles used here, a generous estimate of $\delta T_{Rec}$ is about 30 K.
Due to effects like change in the dielectric constant of the ground with the level of moisture and uncertainties in modeling, $\delta T_{Pick-up}$ is expected to be about $50\%$.
Prior work has estimated the intrinsic error in sky brightness distribution at 408 MHz from \citet{Haslam1982-408-map} to be about $3\%$ \citep{Rogers2004-cal-using-Gbg}.
Scaling $T_{Sky}$ using a spectral index leaves the fractional error unchanged. 
Averaging over a large solid angle patch, as is done here, is expected to reduce it to a lower level.
Due to antenna-to-antenna variations, arising from manufacturing tolerances and imperfections in instrumentation, and the tilt and rotation of the antenna beams due to the gradients in the terrain, the true $P_{N}(\vec{s})$ will differ at some level from the model $P_{N}(\vec{s})$ used here.
These errors in $P_{N}(\vec{s})$ become fractionally larger with increasing angular distance from the beam center. 
Hence, they are less important for the location of the Sun which lies close to the beam center.
These variations have recently been studied in detail by \citet{Neben-2016-MWA_beams}. 
They estimate the net uncertainty due to all the causes considered to be of order $\pm$10--20\% ($1 \sigma$) near the edge of the main-lobe ($\sim$20$^\circ$ from the beam center) and in the side-lobe regions. 
At such distances from the beam center, $P_{N}(\vec{s})$ drops by factors of many to an order of magnitude and can be expected to contribute an independent error of a few percent. 
As the intrinsic uncertainties in the sky model and the $P_{N}(\vec{s})$ cannot be disentangled in our framework, we combine them both in the $\delta \overline{T_{Sky}}$ term and regard it to be about $5\%$.
Given that the geometry of the baseline is known to a high accuracy, $\delta T_{\vec{b}, Sky}$ is primarily due to $\delta \overline{T_{Sky}}$, which is discussed above.
We assume it to scale similarly to $\delta \overline{T_{Sky}}$ and set it to 5\%.

The $\delta S_{\odot,Abs}$, based on the uncertainties discussed above is shown in Fig \ref{Fig:dS_vs_freq}.
To obtain a realistic estimate, the observed value of $\delta r_{N,\odot}$ from Table \ref{Tab:1} was used.
Including the known systematics pushes $\delta S_{\odot,Abs}$ to 10--60\%, for most frequencies, and generally increases with frequency.

\section{Discussion}
\label{Sec:discussion}

\subsection{Flux estimates}
This technique estimates $S_{\odot}$ to lie in the range from $\sim$2--18 SFU in the 100--300 MHz band (Figs. \ref{Fig:3} and \ref{Fig:dS_vs_freq}, and Tables \ref{Tab:1} and \ref{Tab:2}). 
%{\bf 
Being a non-imaging technique, $S_{\odot}$ measures the integrated emission from the entire solar disc.
For the values listed in Tables \ref{Tab:1} and \ref{Tab:2} we use a period which shows a comparatively low level of time variability, or a comparatively quiet time.
These values compare well with earlier measurements \citep[e.g.,][]{McLean-Sheridan-1985}.
The spectrum shown in Fig. 7 peaks at about 240 MHz, in good agreement with theoretical models for thermal solar emission \citep{Martyn1948, Smerd1950, Chambe1978}.
%}
Independent solar flux estimates are available from the Radio Solar Telescope Network (RSTN) at a few fixed frequencies, one of which lies in the range covered here. 
The 245 MHz flux reported by Learmonth station in Australia, in the same interval as presented in Table \ref{Tab:1}, is $\sim$18 SFU. 
The nearest frequency in our data-set is 238 MHz, at which we estimate flux density of 17.1$\pm$0.2 SFU.
Though our measurements are simultaneous they are not at overlapping frequencies.
The period for this comparison was chosen carefully to avoid short-lived emission spikes which often do not have wide enough bandwidth to be seen across even nearby frequencies simultaneously.
A description of the analysis procedure followed at RSTN and the uncertainty associated with these measurements is not available in the literature, though the latter is expected to be about 2--3 SFU (private communication, Stephen White).
Given the associated uncertainties, these measurements are remarkably consistent.

\begin{table}
\caption{Measured values of one of the baselines (Tile011-Tile022).}
\label{Tab:1}
\centerline{\begin{tabular}{cccccccrrcc}
\hline
$\nu$  & $\overline{T_{Sky}}$\tabnote{For $T_{Sky}$ a spectral index of -2.55 is used to translate from 408 MHz to the frequency of observation.} & $T_{\vec{b}, Sky}$ & $T_{Rec}$ & $T_{Pick-up}$ & $\Omega_{P}$ & $r_{N, \odot}$\tabnote{Mean and RMS variation of the measured quantitites are estimated over a period spanning 100 s showing a comparatively low level of time variability.} & $\overline{T_{\odot, P}}^{2}$ & ${S_{\odot}}^{2}$ & $\delta S_{\odot, Th}$ &  
$\overline{T_{\odot}}^{2}$ \\
MHz &  K & K & K & K & sr & & K & SFU & SFU & MK \\
  \hline
103  & 615 & 3.78 & 30 & 20 & 0.380 & 0.174$\pm$0.016 & 139$\pm$019 &  1.73$\pm$0.24 & 0.06 & 0.48$\pm$0.07 \\
117  & 438 & 1.48 & 28 & 17 & 0.340 & 0.273$\pm$0.019 & 184$\pm$013 &  2.64$\pm$0.19 & 0.05 & 0.58$\pm$0.04 \\
131  & 330 & 1.01 & 26 & 15 & 0.283 & 0.384$\pm$0.008 & 236$\pm$007 &  3.53$\pm$0.10 & 0.04 & 0.64$\pm$0.02 \\
148  & 253 & 2.42 & 24 & 13 & 0.236 & 0.510$\pm$0.008 & 307$\pm$008 &  4.86$\pm$0.13 & 0.03 & 0.72$\pm$0.02 \\
167  & 192 & 1.79 & 21 & 12 & 0.220 & 0.593$\pm$0.008 & 336$\pm$010 &  6.30$\pm$0.19 & 0.03 & 0.75$\pm$0.02 \\
189  & 152 & 0.77 & 20 & 12 & 0.240 & 0.623$\pm$0.033 & 323$\pm$045 &  8.47$\pm$1.27 & 0.03 & 0.80$\pm$0.11 \\
213  & 145 & 5.12 & 21 & 13 & 0.228 & 0.653$\pm$0.013 & 356$\pm$023 & 11.30$\pm$0.73 & 0.04 & 0.86$\pm$0.06 \\
240  & 140 & 8.78 & 23 & 18 & 0.202 & 0.709$\pm$0.037 & 498$\pm$119 & 17.77$\pm$4.26 & 0.05 & 1.09$\pm$0.26 \\
272  & 085 & 3.82 & 27 & 10 & 0.150 & 0.702$\pm$0.029 & 345$\pm$047 & 11.77$\pm$1.60 & 0.03 & 0.57$\pm$0.08 \\
299  & 053 & 0.35 & 32 &  9 & 0.128 & 0.696$\pm$0.023 & 277$\pm$035 &  9.73$\pm$1.21 & 0.02 & 0.40$\pm$0.05 \\
  \hline
\end{tabular}}
%\tablenotetext{a}{A spectral index of -2.55 is used to translate from 408 MHz to the frequency of observation.}
%\tablenotetext{b}{Mean and RMS variation over a relatively quiet part of the data spanning 100 s.}
\end{table}

\begin{table}
\caption{Average of values measured for all six baselines.}
\label{Tab:2}
\centerline{\begin{tabular}{ccrrcc}
\hline
$\nu$ & $r_{N, \odot}$\tabnote{Mean and RMS of the measured quantities refer to the mean and RMS for the respective quantities measured over the six baselines considered here.} & $\overline{T_{\odot, P}}^{1}$  & ${S_{\odot}}^{1}$& $\overline{T_{\odot}}^{1}$ &$\theta_{\odot}^{1}$ \\
MHz &  & K & SFU & MK & arcmin \\
  \hline
103  & 0.174$\pm$0.002 &  137.9$\pm$3.3 &  1.71$\pm$0.04 & 0.48$\pm$0.02 & 40.7 \\
117  & 0.270$\pm$0.000 & 181.6$\pm$0.6  &  2.60$\pm$0.01 & 0.58$\pm$0.02 & 40.1 \\
131  & 0.380$\pm$0.000 & 231.4$\pm$0.5  &  3.46$\pm$0.01 & 0.63$\pm$0.02 & 39.5 \\
148  & 0.507$\pm$0.000 & 302.1$\pm$0.6  &  4.79$\pm$0.01 & 0.70$\pm$0.02 & 39.0 \\
167  & 0.592$\pm$0.001 & 270.3$\pm$19.0 &  5.08$\pm$0.35 & 0.63$\pm$0.27 & 38.5 \\
189  & 0.623$\pm$0.001 & 254.8$\pm$18.3 &  6.69$\pm$0.43 & 0.66$\pm$0.27 & 38.1 \\
213  & 0.653$\pm$0.001 & 356.5$\pm$1.2  & 11.30$\pm$0.04 & 0.87$\pm$0.02 & 37.6 \\
240  & 0.703$\pm$0.002 & 480.2$\pm$5.6  & 17.12$\pm$0.20 & 1.06$\pm$0.05 & 37.2 \\
272  & 0.683$\pm$0.031 & 337.4$\pm$27.4 & 11.50$\pm$0.94 & 0.56$\pm$0.01 & 36.9 \\
299  & 0.700$\pm$0.001 & 294.0$\pm$1.9  & 10.30$\pm$0.07 & 0.42$\pm$0.02 & 36.6 \\
  \hline
\end{tabular}}
%\tablenotetext{a}{Mean and RMS refer to the mean and RMS for the respective quantities measured over the six baselines considered here.}
\end{table}

\subsection{Polarization}
The bulk of the data in Fig. \ref{Fig:pol_comparison} clearly follows the x=y line, as is expected for unpolarized thermal emission.
To minimize geometric polarization leakage, the observations presented here were centered at azimuth of $0^{\circ}$.
At the low temperature end, the distribution of data points around the x=y line shows a larger spread, which is reduced at higher temperatures.
The lower temperature measurements come from lower frequencies and this is a manifestation of fractionally larger $\delta S_{\odot, Th}$, or poorer SNR, in this part of the band.
A gradual improvement in the tightness of the distribution along the x=y line is observed as the measured temperature increases and SNR improves with increasing frequency.
Table \ref{Tab:1} shows that data corresponding to quiescent emission lies in the range 50--500 K. 
The data points lying at higher temperatures come from the type III-like event close to the start of the observing period and the numerous fibrils of emission seen at some of the higher frequencies (Fig. \ref{Fig:3}), which are not assured to be unpolarized.
The bulk of the points farther away from the x=y line lie beyond 500 K.
For even the thermal part of the emission, the three highest frequencies do show a systematic departure from the x=y line, which is yet to be understood.
Similar behavior is seen for other baselines which were studied.

\subsection{Comparison across baselines}
To build a quantitative sense for the uncertainty in the estimates of $T_{\odot}$, we examine the ratios of $T_{\odot,P}$ measured on different baselines (Figs. \ref{Fig:baseline_comparison1} and \ref{Fig:baseline_comparison2}).
The histograms of this ratio are symmetric and Gaussian-like in appearance. 
At the frequency band where the widest spread is seen (167 MHz), the medians of these histograms lie in the range 0.8--1.2. 
For many of the frequencies this spread is larger than the expectations based on the widths of these Gaussians.
The ratios of baseline pairs exhibit smooth trends, as opposed to showing random fluctuations. 
The observed spread is likely due to the systematic effects which give rise to antenna-to-antenna differences, and were not accounted for in the analysis.
Table \ref{Tab:2} lists the mean and rms values of various quantities of interest over all six baselines.
The RMS in the estimates of $S_{\odot}$ due to these systematic effects across many baselines (Table \ref{Tab:2}) is usually less than that due to the intrinsic variations in $S_{\odot}$ on a given baseline (Table \ref{Tab:1}).
It is usually a few percent or less and the largest value observed is about 8\%.

\subsection{Uncertainty analysis}
An analysis of various known sources of systematic errors lead to an uncertainty in the absolute values of $S_{\odot}$ estimated not exceeding $\pm 60\%$, except at 240 MHz, where the value is larger due to the much larger variation in $\delta r_{N, \odot}$ (Fig. \ref{Fig:3}, Table \ref{Tab:1} and Sec. \ref{Sec:errors}).
The observed baseline-to-baseline variation is significantly smaller than this uncertainty estimate (Fig. \ref{Fig:dS_vs_freq}).
This suggests that at least some of the values for uncertainties in the individual parameters of the system considered in Sec. \ref{Sec:errors} are over-estimates.
We also note that the uncertainty in relative values of $S_{\odot}$ from a given interferometric baseline, which is determined primarily by the $\delta S_{\odot, Th}$, is much smaller than that in its absolute value.
These measurements, hence have the ability to quantify variations in the observed values of $S_{\odot}$, as small as about a percent.

\section{Conclusions}
\label{Sec:conclusions} 
We have demonstrated that this technique provides a convenient and robust approach for determining the solar flux using radio interferometric observations from a handful of suitable baselines.
It is much less intensive in terms of the data ($<$0.1\% in the case of MWA), and the human and computational effort it requires, when compared to conventional interferometric analysis.
It provides flux estimates with fair absolute accuracy and can reliably measure relative changes of order a percent.
As it provides flux estimates at the native resolution of the data, this technique is equally applicable for quiet and active solar emissions.
Further, on assuming an angular size for the Sun, this approach can also provide the average brightness temperature of the Sun.

Good solar science requires monitoring-type observations, where a given instrument observes the Sun for the longest duration feasible every day of the year.
However, given the enormous rate at which data is generated by the new technology low radio frequency interferometers and the requirements of solar imaging to maintain high time and spectral resolution in the image domain, it is currently not possible for the conventional analysis methods to keep up with the rate of data generation.
Efficient techniques and algorithms need to be developed not only to image these data, but also to synthesize the information made available in the 4D image cubes in a meaningful humanly understandable form.
In the interim, algorithms like the one presented here enable some of the novel and interesting science made accessible by these data. 
For wide-band low radio frequency observations, which are increasingly becoming more common, this technique will allow simultaneous characterization of the coronal emissions over a large range of coronal heights.
This technique can naturally be implemented for other existing and planned wide field-of-view instruments with similar levels of characterization, like LOFAR, LWA and SKA-Low.
Amenability of this technique to automation will enable studies involving large volumes of data, which will, in turn, open the doors to multiple novel investigations addressing fundamental questions ranging from variations in the solar flux as a function of solar cycle to quantifying the short-lived narrow-band weak emission features seen in the wideband low radio frequency solar data and exploring their role in coronal heating.  
 
\begin{acks}
We acknowledge Randall Wayth and Budi Juswardy, both at Curtin University, Australia, for helpful discussions and providing estimates of $T_{Rec}$ and $\delta T_{Rec}$. 
We also acknowledge helpful comments from Stephen White (Air Force Research Laboratory, Kirtland, NM, USA) and David Webb (Boston College, MA, USA) on an earlier version of the manuscript.
This scientific work makes use of the Murchison Radio-astronomy Observatory, operated by CSIRO. We acknowledge the Wajarri Yamatji people as the traditional owners of the Observatory site.  Support for the operation of the MWA is provided by the Australian Government Department of Industry and Science and Department of Education (National Collaborative Research Infrastructure Strategy: NCRIS), under a contract to Curtin University administered by Astronomy Australia Limited. We acknowledge the iVEC Petabyte Data Store and the Initiative in Innovative Computing and the CUDA Center for Excellence sponsored by NVIDIA at Harvard University.
{\it Facilities:} \rm{Murchison Widefield Array}.
\end{acks}

\end{article} 

\end{document}